\documentclass[aps,prl,twocolumn,showpacs,floatfix]{revtex4}
\usepackage{epsfig}
\usepackage{graphicx}
\usepackage{dcolumn}
\usepackage{amsthm,amsmath}

\begin{document}

\title{Enhanced spin-dependent parity non-conservation effect in the $7s \ ^2S_{1/2} \rightarrow 6d \ ^2D_{5/2}$ transition in Fr: 
A possibility for unambiguous detection of nuclear anapole moment}
\vspace*{0.5cm}

\author{$^a$B. K. Sahoo \footnote{bijaya@prl.res.in}, $^b$T. Aoki, $^c$B. P. Das and $^d$Y. Sakemi}
\affiliation{$^a$Theoretical Physics Division, Physical Research Laboratory, Ahmedabad-380009, India\\
$^b$Institute of Physics, Graduate School of Arts and Sciences, The University of Tokyo, 153-8902, Japan\\
$^c$International Education and Research Center of Science and Department of Physics, Tokyo Institute of Technology, 2-1-2-1-H86 Ookayama
Meguro-ku, Tokyo 152-8550, Japan\\
$^d$Cyclotron and Radioisotope Center, Tohoku University, Sendai, Miyagi 980-8578, Japan}

\date{Recieved date; Accepted date}
\vskip1.0cm

\begin{abstract}
\noindent
Employing the relativistic coupled-cluster method, comparative studies of the parity non-conserving electric dipole amplitudes
for the $7s \ ^2S_{1/2} \rightarrow 6d \ ^2D_{5/2}$ transitions in $^{210}$Fr and $^{211}$Fr isotopes have been carried out. 
It is found that these transition amplitudes, sensitive only to the nuclear spin dependent effects, are enhanced by more 
than 3 orders compared to the low-lying $S-D_{5/2}$ transitions in Ba$^+$ and Ra$^+$ owing to the very large contributions 
from the electron core-polarization effects in Fr. This translates to a relatively large and, in principle, measurable induced 
light shift, which would be a signature of nuclear spin dependent parity nonconservation that is dominated by the nuclear anapole 
moment in a heavy atom like Fr. A plausible scheme to measure this quantity using the Cyclotron and Radioisotope Center (CYRIC) 
facility at Tohoku University has been outlined.
\end{abstract}

\maketitle

The study of parity nonconservation (PNC) effects in atomic systems, which involve the interplay between the weak and electromagnetic interactions 
\cite{bouchiat}, has important implications for atomic physics, nuclear physics and particle physics \cite{commins,flambaum, erler}. For 
example, it could (i) provide hints for the possible existence of new physics beyond the standard model (SM) of particle interactions 
\cite{marciano}, (ii) probe the existence of the nuclear anapole moment (NAM) which is presumed to be a fundamental property of an atomic 
nucleus \cite{flambaum, ginges}, and (iii) test the role of the electron correlation effects in a parity nonconserving electric dipole 
transition amplitude that depends on the region near and far away from the nucleus \cite{bijaya1}. A {high precision }PNC measurement for 
the $6s \ ^2S_{1/2} \rightarrow 7s \ ^2S_{1/2}$ transition in Cs has yielded a result that is in good  agreement with the SM  
\cite{wood, porsev}, and it has also led to the observation of its NAM with an accuracy of 15\% \cite{wood}. However, it is at variance with the 
results of the shell model and the nucleon-nucleon scattering experiments \cite{wilburn,haxton}. It is, therefore, imperative to  search for 
NAMs in other systems. Because of this reason, a PNC measurement was carried out on the $6s^2 \ ^1S_0 \rightarrow 5d6s \ ^3D_1$ transition 
in Yb \cite{tsigutkin}. It is indeed desirable to observe an NAM unambiguously in an atomic system. N. Fortson has made an important proposal 
to measure PNC using a single trapped ion \cite{fortson} based on the observation of the PNC induced light shift, which arises due to the 
interference of the parity nonconserving electric dipole and the electric quadrupole (E2) amplitudes of a transition such as the $6s \ ^2S_{1/2} 
\rightarrow 5d \ ^2D_{3/2}$ transition in Ba$^+$. Though the choice of a single ion would limit  the statistical uncertainty, it can be partly 
compensated by selecting a transition such that the upper state has a long lifetime like the above transition in Ba$^+$ \cite{fortson}, and 
furthermore, the large storage time in a trap contributes to enhancing the sensitivity of the scheme proposed by Fortson. As a consequence, the 
forbidden low-lying $S-D$ transitions in the singly charged Ba \cite{fortson}, Yb \cite{bijaya2} and Ra \cite{wansbeek} ions have been considered 
for the PNC studies. In fact, it has also been pointed out that existence of the NAM can be unambiguously inferred from the measurements of the 
nuclear spin dependent (NSD) PNC in the $S-D_{5/2}$ transitions of these ions using the techniques similar to the observation of the light shift 
techniques of Fortson \cite{geetha,bijaya4}. The major disadvantage of these transitions is that their $E1^{PNC}$ amplitudes are small 
\cite{geetha,bijaya4,roberts}. The measurement of the NAM  of Fr has been proposed in \cite{gomez1,sheng} by considering the hyperfine transitions 
of the ground state of that atom. In this Letter, we demonstrate that the $E1^{PNC}$ amplitude for the $7s \ ^2S_{1/2} \rightarrow 6d \ ^2D_{5/2}$ 
transition in Fr, which arises only from the NSD interaction, is enhanced relative to the same transition in the heavy ions mentioned above. Thus, 
the PNC light shift for this transition will also be enhanced. So if it can be measured successfully then it would constitute an unambiguous 
signature of the NAM.

\begin{table}[t] \small
\caption{Reduced matrix elements $\mathcal{Y}$ in $iea_0 {\cal K}_W \times 10^{-11}$ from different calculations. We only mention magnitudes 
from the other references: $^a$\cite{johnson} and $^b$\cite{roberts}.}
\begin{ruledtabular}
\begin{tabular}{lcccc}
 $J_f \rightarrow J_i$  & $F_f$  &  $F_i$ &  This work  & Others \\ 
\hline
\multicolumn{5}{l}{\underline{$^{210}$Fr ($I=6$)}} \\
$7s \ ^2S_{1/2} \rightarrow 7s  \ ^2S_{1/2}$   &  11/2   &  13/2   &  37.39 & \\
$8s \ ^2S_{1/2} \rightarrow 7s  \ ^2S_{1/2}$   &  11/2   &  11/2   & $-6.28$  \\
                                               &  13/2   &  11/2   & $-15.70$  \\
                                               &  11/2   &  13/2   & $-17.13$  \\
                                               &  13/2   &  13/2   & $-6.69$   \\
$6d  \ ^2D_{3/2} \rightarrow 7s  \ ^2S_{1/2}$  &  9/2   &  11/2   &  $-20.90$    \\
                                               &  11/2   &  11/2   &  22.16 \\
                                               &  13/2   &  11/2   &  $-20.09$     \\
                                               &  11/2   &  13/2   &   14.82  \\
                                               &  13/2   &  13/2   &   $-20.38$   \\
                                               &  15/2   &  13/2   &  22.67 \\
$6d  \ ^2D_{5/2} \rightarrow 7s  \ ^2S_{1/2}$  &  9/2   &  11/2   &  7.70  \\
                                               &  11/2   &  11/2   &  $-9.55$    \\
                                               &  13/2   &  11/2   &  8.17  \\
                                               &  11/2   &  13/2   &  5.98  \\
                                               &  13/2   &  13/2   &   $-10.34$   \\
                                               &  15/2   &  13/2   &  12.39 \\
\multicolumn{5}{l}{\underline{$^{211}$Fr ($I=9/2$)}} \\
$7s \ ^2S_{1/2} \rightarrow 7s  \ ^2S_{1/2}$   &  4.0   & 5.0    &  22.47 & 23.79$^a$ \\
$8s  \ ^2S_{1/2} \rightarrow 7s  \ ^2S_{1/2}$  &  4.0   & 4.0    & $-4.15$ & 3.092$^a$ \\
                                               &  5.0   & 4.0    & $-10.43$ & 9.224$^a$ \\
                                               &  4.0   & 5.0    & $-11.67$ & 10.16$^a$  \\
                                               &  5.0   & 5.0    & $-4.59$  & 3.426$^a$ \\                                            
$6d  \ ^2D_{3/2} \rightarrow 7s  \ ^2S_{1/2}$  & 3.0    &  4.0   & $-13.75$   \\
                                               & 4.0    &  4.0   &  15.11  \\
                                               & 5.0    &  4.0   & $-14.11$ \\
                                               & 4.0    &  5.0   & 9.49  \\
                                               & 5.0    &  5.0   & $-13.54$  \\
                                               & 6.0    &  5.0   & 15.33 \\                            
$6d  \ ^2D_{5/2} \rightarrow 7s  \ ^2S_{1/2}$  & 3.0    &  4.0   & 4.66 & 0.243$^b$ \\
                                               & 4.0    &  4.0   & $-6.25$  & 0.326$^b$ \\
                                               & 5.0    &  4.0   & 5.68 & 0.296$^b$ \\
                                               & 4.0    &  5.0   & 3.77 & 0.197$^b$ \\
                                               & 5.0    &  5.0   & $-6.96$ & 0.363$^b$ \\
                                               & 6.0    &  5.0   & 8.78 & 0.458$^b$   \\
\end{tabular}
\end{ruledtabular}
\label{tab1}
 \end{table}
 
We employ the relativistic coupled-cluster (RCC) theory in the singles, doubles and important partial triples excitations approximation 
(CCSD$_{\text{t3}}$ method) \cite{bijaya6,bijaya7} to evaluate the NSD PNC amplitudes corresponding to the transitions between different hyperfine 
levels of the ground state and also the $7s \ ^2S_{1/2} \rightarrow 8s \ ^2S_{1/2}$, $7s \ ^2S_{1/2} \rightarrow 6d \ ^2D_{3/2}$ and $7s \ ^2S_{1/2} \rightarrow 
6d \ ^2D_{5/2}$ transitions of $^{210}$Fr and $^{211}$Fr. The principle of the experiment to observe their signature is given later, and it involves the measurement of the PNC induced light shift of the 
$7s \ ^2S_{1/2} \rightarrow 6d \ ^2D_{5/2}$ transition in $^{210}$Fr. In addition, we also present results for the other three transitions and for those corresponding to
$^{211}$Fr isotope for three specific reasons. Firstly, the calculations of the NSD PNC amplitudes in some of these transitions for $^{211}$Fr 
have already been reported using different relativistic many-body methods \cite{roberts,johnson} and it is instructive to compare their  results with those obtained by us. Secondly, our calculations for $^{211}$Fr could be useful for another experiment 
involving ground  state hyperfine transitions (e.g. \cite{gomez1,sheng}). Thirdly, the nuclear spin $I$ of $^{210}$Fr and $^{211}$Fr are integer ($I=6$) 
and half-integer ($I=9/2$), respectively, which can be appropriately used in different experimental set-ups.
 
The Hamiltonian due to the NSD PNC interaction is given by \cite{ginges}
\begin{eqnarray}
H_{PNC}^{NSD} &=& \frac {G_F}{\sqrt{2}} {\cal K}_W \mbox{\boldmath$\alpha$} \cdot {\bf I} \ \rho_{nuc}(r) ,
\label{eq2}
\end{eqnarray}
where $G_F$ is the Fermi constant, $\rho_{nuc}$ is the nuclear density and $\mbox{\boldmath$\alpha$}$ is the Dirac matrix. In the above expression
the dimensionless quantity ${\cal K}_W$ is related to NAM. The $E1^{PNC}$ amplitude due to the NSD interaction between the hyperfine states 
$|F_f,M_f \rangle$ and $|F_i,M_i \rangle$ is given by
\begin{eqnarray}
E1^{PNC}_{M_f M_i} &=& (-1)^{F_f - M_f} \left ( \begin{matrix} F_f & 1 & F_i \cr -M_f & q & M_i \cr \end{matrix} \right ) \mathcal{Y},
\label{eq4}
\end{eqnarray}
where $q=-1$, 0 or 1 depends on the choice of the $M$-values. For the theoretical purpose, enhancement in the $E1^{PNC}$ is estimated by 
calculating the reduced matrix element $\mathcal{Y}$ given by
\begin{eqnarray}
\mathcal{Y} &=& \eta \ \Big ( \sum_{k\ne i} (-1)^{j_i - j_f +1} \frac{ \langle J_f || D || J_k \rangle \langle J_k || K^1 || J_i \rangle}{ E_i - E_k} \nonumber \\ 
&& \times  \left \{ \begin{matrix} F_f & F_i & 1 \cr J_k & J_f & I \cr \end{matrix} \right \} \left \{ \begin{matrix} I & I & 1 \cr J_k & J_i & F_i \end{matrix} \right \} \nonumber \\ 
&& + \sum_{k\ne f} (-1)^{F_i - F_f +1} \frac{ \langle J_f || K^1 || J_k \rangle \langle J_k || D || J_i \rangle}{ E_f - E_k} \nonumber \\
&&  \times \left \{ \begin{matrix} F_f & F_i & 1 \cr J_i & J_k & I \cr \end{matrix} \right \} \left \{ \begin{matrix} I & I & 1 \cr J_k & J_f & F_f \cr \end{matrix} \right \} \Big ),
\label{eq5}
\end{eqnarray}
where $\eta=\sqrt{I(I+1)(2I+1) (2F_i+1) (2F_f+1)}$ and $E$s are the energies of the respective states. The above expression is derived by
rewriting $H_{PNC}^{NSD}=\sum_q (-1)^q I_q K^q$. We determine these quantities in a sum-over-states approach by calculating the reduced 
matrix elements of the $D$ and $K$ operators using the CCSD$_{\text{t3}}$ method. In this method, the wave function ($|\Psi_n \rangle$) of 
an atomic state corresponding to the closed-shell configuration $[6p^6]$ and a valence orbital $n$ in Fr is expressed as 
\begin{eqnarray}
  |\Psi_n \rangle = e^{T_1+T_2} \{1+S_{1n}+S_{2n} \} |\Phi_n \rangle,
\end{eqnarray}
where $T$ and $S_n$ are the excitation operators involving the core and core-valence electrons, respectively, with subscripts 1 and 2 
representing the levels of excitations. The reference state $|\Phi_n \rangle$ is obtained by $|\Phi_n \rangle=a_n^{\dagger} |\Phi_0 \rangle$, 
where $|\Phi_0 \rangle$ is the Dirac-Fock (DF) wave function of the closed-core $[6p^6]$. This method has already been applied earlier to evaluate 
both the hyperfine structure constants and radiative transition matrix elements of the Fr isotopes accurately \cite{bijaya6,bijaya7}. We, 
however, include contributions explicitly only from the $7P-11P$ low-lying states obtained using the CCSD$_{\text{t3}}$ method while the 
other smaller contributions such as from the core-valence, higher level excitations etc. are estimated using the second order many-body 
perturbation theory (MBPT(2) method). 

\begin{table}[t]
\caption{Estimated light shifts in the hyperfine levels of the $7s \ ^2S_{1/2}(F_i) \rightarrow 6d \ ^2D_{5/2} (F_f)$ transition 
of $^{210}$Fr due to the E2 (in MHz) and NSD PNC (in $\times 10^{-4}$ Hz) interactions with the applied electric field $2 \times 10^6 \ V/m$.  
Here we have used ${\cal K}_W \approx 0.57$ and E2 amplitude as 39.33 $e^2a_0^2$.}
\begin{ruledtabular}
\begin{tabular}{lcccc}
 $F_f$  &  $F_i$ & $M$ & $\Delta \omega^{E2}_{|M|}/2\pi$  & $\Delta \omega^{PNC}_M/2\pi$ \\ 
\hline
 & & \\
  9/2    &  11/2   &  1/2  & 9.15  & $-51.4$  \\
 11/2    &  11/2   &  1/2  & 2.32  & $1406.7$ \\
 13/2    &  11/2   &  1/2  & 6.41  & $49.6$ \\ 
 11/2    &  13/2   &  1/2  & 5.46 & $26.8$ \\
 13/2    &  13/2   &  1/2  & 1.70  & $-1414.1$ \\
 15/2    &  13/2   &  1/2  & 7.91  & $-51.3$ \\
\end{tabular}
\end{ruledtabular}
\label{tab2}
 \end{table}
 
 In Table \ref{tab1}, we give the results of the reduced matrix elements $\mathcal{Y}$ from our calculations of the previously mentioned 
hyperfine transitions of $^{210}$Fr and $^{211}$Fr. We also compare our results with the other available calculations for $^{211}$Fr 
\cite{roberts,johnson}. All these calculations start with a $V^{N-1}$ potential, but Johnson et al have employed the random phase approximation 
(RPA) to calculate the $\mathcal{Y}$ values only for the $S-S$ transitions \cite{johnson}. Our CCSD$_{\text{t3}}$ method contains these effects 
implicitly along with the core-correlation and pair-correlation effects to all orders. Nevertheless our calculations agree quite well with these 
RPA results, the differences mainly owing to the pair-correlation effects that are significant for the $S$ states as seen in the studies of the
hyperfine structure constants of $^{210}$Fr \cite{bijaya7}. Recently, Roberts et al have calculated $\mathcal{Y}$ values for the $7s \ ^2S_{1/2} \rightarrow 
6d \ ^2D_{5/2}$ transition that take into account the core-valence correlation effects using the correlation potential (CP) method and the 
polarization of the core electrons and interactions with the external fields using RPA. But they mention that their results can be improved after 
inclusion of the other higher order correlation corrections such as the double-core-polarization, structural radiation and ladder-diagrams \cite{roberts}. 
On comparison, we find at least one order of magnitude difference between their results and ours. Our analysis shows that the extraordinarily large 
core-polarization effects enhance the $\langle 6d \ ^2D_{5/2} | K^1 | np \ ^2P_{3/2} \rangle$ matrix elements that appear in the second term of 
Eq. (\ref{eq5}), and their values become more than two times larger than the  $\langle np \ ^2P_{3/2} | K^1 | 7s \ ^2S_{1/2} \rangle$ matrix elements, 
where $n$ represents the principal quantum numbers of $p$ orbitals. The two factors that are responsible for such enhancements are the small 
energy difference between the $6d \ ^2D_{5/2}$ and $7p \ ^2P_{3/2}$ states and large overlap between the valence $7s$ orbital and the 
occupied $p_{1/2}$ orbitals. The other factors that also play vital roles here are the large $\langle 6d \ ^2D_{5/2} | D | 7p \ ^2P_{3/2} 
\rangle$ matrix element ($\sim$10.51(7) $ea_0$ \cite{bijaya6}) and the positioning of the the $7p \ ^2P_{3/2}$ state in between the 
$7s \ ^2S_{1/2}$ and $6d \ ^2D_{5/2}$ states. Due to the same reason, the enhancement for this transition in Fr is much larger than  
its isoelectronic partner Ra$^+$ and also for the $6s \ ^2S_{1/2} \rightarrow 5d \ ^2D_{5/2}$ transition of Ba$^+$ \cite{bijaya4}. It is also 
observed that $\mathcal{Y}$ values in $^{210}$Fr are larger than $^{211}$Fr due to its large $I$.

\begin{figure}[t]
\begin{center}
\includegraphics[width=8.5cm, height=8.5cm, clip=true]{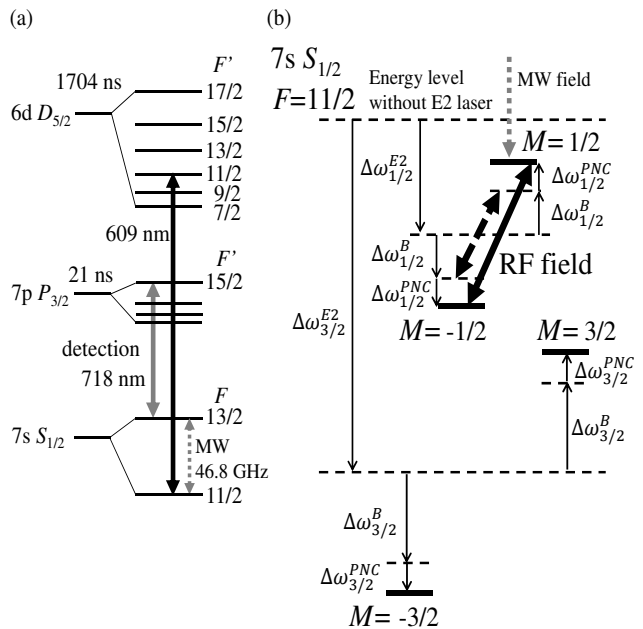}
\label{fig1}
\end{center}
\caption{Schematic energy level diagrams of $^{210}$Fr. (a) Arrows indicate laser induced transitions for observing the E2 light shifts, detecting the
states, and carrying out the micro-wave (MW) transitions between the hyperfine levels. (b) Magnetic sublevels (shown only for $M=\pm 3/2$ and 
$M=\pm 1/2$) of the $F=11/2$ level of the $7s \ ^2S_{1/2}$ state with the corresponding RF transitions. The solid and dashed arrows indicate the 
resonant RF transitions in the presence and absence of the PNC induced light shift, respectively.} 
\label{fig1}
\end{figure}
We suggest an approach similar to that proposed by Fortson \cite{fortson} to measure the NSD PNC induced light shift  ($\omega^{PNC}$) 
arising from the interference of the NSD $E1^{PNC}$ and E2 amplitudes between the hyperfine states of the $7s \ ^2S_{1/2} \rightarrow 6d \ 
^2D_{5/2}$ transition in $^{210}$Fr. Fig. \ref{fig1} shows schematic diagrams of the relevant transitions for the PNC measurement and also  
indicates that the $7s \ ^2S_{1/2} \rightarrow 6d \ ^2D_{5/2}$ transition is in the optical regime. The frequency for a transition with the 
same hyperfine sub-levels $M$ can be estimated using the expression \cite{fortson}
\begin{eqnarray}
\label{eqn1}
\Delta\omega_M^{PNC} &\approx& - \frac{Re \sum_{M'}(\Omega_{M M'}^{PNC*}\Omega_{M M'}^{E2})} {\sqrt{\sum_{M'}|\Omega_{M M'}^{E2}|^2}},
\end{eqnarray}
where $\Omega^{PNC}$ and $\Omega^{E2}$ are the Rabi frequencies due to the $E1^{PNC}$ and E2 amplitudes and the summation over $M'$ is for all
possible allowed intermediate states. This will be much smaller compared to the changes in the transition frequency due to the E2 shift 
alone, which is given by
\begin{eqnarray}
\Delta\omega_{M}^{E2} &\approx& \frac{(\omega_0 - \omega)}{2} - \sqrt{\sum_{M'}|\Omega_{M M'}^{E2}|^2}
\label{eqn2}
\end{eqnarray}
for the respective frequencies $\omega_0$ and $\omega$ corresponding to the transition before and after applying the laser. In the nuclear 
shell model, $^{210}$Fr has an odd proton in the $\pi h_{9/2}$ shell and an odd neutron in the $\nu f_{5/2}$ shell. We determine ${\cal K}_W$
of this isotope considering the dominant contribution from the odd proton due to the NAM using the expression \cite{flambaum}
\begin{eqnarray}
 {\cal K}_W \approx \frac{9}{10} g_p \mu_p \frac{\alpha A^{2/3}}{M_p r_0},
\end{eqnarray}
where $g_p \simeq 5.0$ is the gyromagnetic factor and $\mu_p \simeq 2.8$ is the magnitude of the magnetic moment of the proton, $A$ is the 
atomic number, $M_p$ is the proton mass and $r_0 \simeq 1.2$ fm. Considering the $\mathcal{Y}$ values from Table \ref{tab1}, $M=1/2$, 
${\cal K}_W \simeq 0.57$ from the above formula, the values of the electric field and the E2 amplitude are $2 \times 10^6 \ V/m$ and 39.33 
$e^2a_0^2$ \cite{bijaya6} respectively, we have estimated $\Delta\omega_{M}^{E2}$ and $\Delta\omega_M^{PNC}$ values for different hyperfine levels 
of the $7s \ ^2S_{1/2} \rightarrow 6d \ ^2D_{5/2}$ transition in $^{210}$Fr and presented in Table \ref{tab2}. We find that there are significant 
enhancements in the PNC induced light shifts for the $F_i=11/2 \rightarrow F_f=11/2$ and $F_i=13/2 \rightarrow F_f=13/2$ transitions. The 
measurements of these quantities are possible using the laser cooled $^{210}$Fr atoms in an optical lattice that is being set-up at the Cyclotron and 
Radioisotope Center (CYRIC), Tohoku University. We plan to irradiate two standing-wave laser fields (similar to that suggested in Ref. \cite{fortson}) 
with wavelengths of 609 nm, which are resonant with the $7s \ ^2S_{1/2} (F=11/2) \rightarrow 6d \ ^2D_{5/2} (F^{\prime}=11/2)$ E2 transition, as 
shown in the black solid arrow in Fig. \ref{fig1}(a). The PNC induced light shift, which is given by Eq. (\ref{eqn1}), can be measured from the 
Ramsey resonance by applying the radio frequency (RF) field, as shown in Fig. \ref{fig1}(b), arising from two-pulses separated in time. Thus, the 
magnetic sublevels $M = \pm 1/2$ of the $7s \ ^2S_{1/2}(F=11/2)$ state can be shifted in this scheme by the E2 light field ($\Delta 
\omega^{E2}_{|M|}$), Zeeman effect ($\Delta \omega^{B}_{|M|}$) and the PNC induced effect ($\Delta \omega^{PNC}_{M}$); where $|M|$ in the subscripts 
indicate that the corresponding shifts are independent of the sign of $M$. We estimate the values of $\Delta \omega^{E2}_{|M|}$ for $M=\pm 1/2$ and 
$\pm 3/2$ as $\sim2.31$ MHz and $\sim5.00$ MHz respectively. Similarly, $\Delta \omega^{B}_{1/2} / 2\pi$ is $\sim0.1$ MHz for a magnetic field 
strength of 1 G, which is much smaller than the estimated E2 light shifts. The frequency of the RF field will be resonant only with the $M = -1/2
\leftrightarrow M=1/2$ transition. As shown in Fig. \ref{fig1}(b), this RF field will not induce a transition between any other magnetic sublevels.
Interactions of the two-pulses with the $^{210}$Fr atoms can produce the Ramsey fringes. After the application of the two-pulses of the RF field, 
the states selective detection can be achieved from the following manner: (i) A microwave (MW) field of 46.8 GHz is applied, causing a transition 
from the $|F=11/2,M=1/2 \rangle$ level to the $|F=13/2, M=1/2 \rangle$ level, as shown in Fig. \ref{fig1}(a) and (b). (ii) Using laser to drive the 
transition from the $7s \ ^2S_{1/2} (F=13/2)$ state to the $7p \ ^2P_{3/2} (F^{\prime}=15/2)$ state (718 nm optical transition) and detecting the 
fluorescence, representing the Ramsey fringes from this transition, can be detected by a photo multiplier tube in order to measure the final 
population in the $M=1/2$ state. The RF transition mentioned earlier will not be affected by the E2 light shift fluctuation due to the amplitude 
noise of the laser fields but by the PNC light shifts owing to opposite signs for the $\Delta \omega^{PNC}_{M}$ values for the $M = -1/2$ and $M=1/2$ 
levels. Therefore, comparing the phase shifts of the Ramsey fringes in the presence and absence of the E2 laser field would yield a net shift equal 
to $2 \Delta \omega^{PNC}_{1/2}$. The uncertainty in the measurement would be restricted by the shot noise limit given by $\delta \omega / 2\pi 
= 1/(2\pi \sqrt{\tau N T})$ \cite{fortson}, where $\tau$ is the separation time between the two RF pulses, $N$ is the number of trapped atoms, and $T$ is the total measurement time. If $\tau$ is 
$\sim 1704$ns, the lifetime of the $6d \ ^2D_{5/2}$ state \cite{bijaya6}, and $N = 10^4$ - the number of atoms that is expected to be trapped in 
CYRIC, $T$ should be more than 20 s in order to obtain the PNC induced light shift of 0.28 Hz.

In summary, we have analyzed the PNC induced light shifts in the $7s \ ^2S_{1/2} \rightarrow 6d \ ^2D_{5/2}$ transition in $^{210}$Fr and have
proposed a plausible experimental scheme to measure them using the CYRIC facility. Calculations in $^{210}$Fr and $^{211}$Fr using the RCC theory 
showed overwhelmingly large enhancements of the PNC amplitudes; more than 3 orders of magnitude compared with those of Ba$^+$ and Ra$^+$,
due to strong core-polarization effects. This suggests that an unambiguous observation of the NAM in Fr is possible. Our calculations of the PNC 
amplitudes for $^{211}$Fr could be useful if the corresponding PNC induced light shift measurements are carried out in another facility. 

We would like to thank Dr. M. Mukherjee for many useful discussions on light shifts. BKS thanks Professor V. V. Flambaum for explaining the 
anapole moment expression. This work was supported partly by INSA-JSPS under project no. IA/INSA-JSPS Project/2013-2016/February 28,2013/4098. 
Computations were carried out using the PRL Vikram-100 HPC cluster.

\end{document}